\documentstyle[preprint,aps]{revtex}
\begin{document}
\title{Dynamic and geometric alignment of CS$_2$ in 
intense laser fields of picosecond and femtosecond duration}  
\author{S. Banerjee, G. Ravindra Kumar, and D. Mathur} 
\address{
Tata Institute of Fundamental Research, Homi Bhabha Road,
Mumbai 400 005, India.
}
\date{\today}
\maketitle
\begin{abstract}
CS$_2$ is identified as a molecule for which distinction can 
be made between dynamic and geometric alignment induced
by intense laser fields. Measured anisotropic angular 
distributions of fragment ions arise from (i) dynamic alignment 
of the S-C-S axes along the laser polarization vector for 
35-ps laser pulses and (ii) geometric alignment due to 
an angle-dependent ionization rate in the case of 
100-fs pulses. Results of classical calculations of the 
alignment dynamics support our observations.
By comparing mass spectra obtained with linearly- 
and circularly-polarized light it is not possible to distinguish 
between dynamic and geometric alignment. 
\end{abstract}


Prevailing wisdom on the spatial alignment of molecules 
in intense
laser fields has been challenged in two recent 
reports \cite{posthumus,ellert}. The possibility of spatially 
aligning molecules using strong light has attracted much
attention since the pioneering double-pulse experiments of
Normand {\it et al.} and Dietrich {\it et al.} \cite{normand} 
appeared to firmly establish that the field associated
with intense, linearly-polarized, laser light of picosecond 
duration induces sufficiently strong
torques on an initially-randomly oriented ensemble of 
linear diatomic molecules for reorientation of 
internuclear axis to occur. Experimental 
manifestation of such alignment is the anisotropic 
angular distribution of fragments produced upon 
subsequent dissociative ionization of
molecules: ion intensities are maximum in the 
direction of the laser polarization vector and minimum 
(frequently zero) in the orthogonal direction. Earlier 
work on diatomic molecules has been extended to 
triatomics and polyatomics \cite{ourwork}and recently for neutrals \cite{sakai};
much of the interest in the  
alignment of the internuclear axis ( referred to in the current literature 
as dynamic alignment) in molecules has been generated 
because of tantalizing possibilities of  pendular-state 
spectroscopy \cite{pendular} and coherent control 
experiments \cite{charron}. Now, results of experiments 
conducted by Posthumus {\it et al.}  \cite{posthumus} 
and by Ellert and Corkum \cite{ellert} offer indications 
that when a linearly-polarized light field acts on 
molecules whose constituent atoms are heavy 
(such as iodine-containing diatomics and polyatomics), 
laser-induced dynamic alignment may not occur. The
angular distributions of the products of dissociative 
ionization in such cases might be determined 
essentially by the dependence of the
ionization rate on the angle made by the laser polarization
(conventionally referred to as photoionization anisotropy)
vector with the molecule's symmetry axis 
\cite{posthumus,ellert}. For a given value of laser 
intensity, the rate of ionization is largest for those 
molecules whose internuclear axis lies parallel to the 
direction of the laser polarization vector. The observed 
anisotropy of the fragment ion angular distribution is 
therefore determined by a purely geometric 
ffect - namely the angle made by the molecule with 
the light field direction.  Moreover, this 
increase in ionization rate maximizes at a critical 
internuclear distance at which the least bound electron 
localizes on one atomic core and the field of the other 
core adds to the laser field, causing the elongated
molecule to field ionize. Posthumus {\it et al.} 
\cite{posthumus} have presented a classical model for 
such enhanced ionization in which
it is not necessary to invoke dynamic alignment 
in order to account for anisotropic angular distributions 
of the products of laser-field-induced dissociative ionization 
of I$_2$. In the light of these developments, it is clearly 
important to reassess the contribution of dynamic and 
geometric alignment to the observed anisotropy of 
angular distributions of fragment ions, especially 
when femtosecond duration light pulses are used to 
field ionize molecules.  

Proper theoretical insight into the extent of alignment 
obtained when molecules are irradiated by intense 
laser fields is difficult to attain due, essentially, to 
unknown values of polarizabilities and
hyperpolarizabilities of the gamut of electronic states that
might be accessed in the course of dissociative ionization. 
As noted by Ellert and Corkum \cite{ellert}, it is 
of much importance to experimentally assess the 
extent to which the angular anisotropies measured in 
earlier picosecond and femtosecond 
experiments are actually due to dynamic alignment. 
To this end we report here the results of experiments 
on the linear triatomic, CS$_2$, using 
picosecond laser beams (35 ps, 532 nm) in
the intensity range 10$^{13}$ W cm$^{-2}$ and 
femtosecond beams (100 fs, 806 nm) in the range 
10$^{13}$ W cm$^{-2}$-
10$^{15}$ W cm$^{-2}$. CS$_2$, along with its valence
isoelectronic companion, CO$_2$,  is the archetypal 
triatomic system that has been subjected to many 
experimental studies. In the context of
the present work, it also represents a species on the 
boundary between `heavy' molecules (such as I$_2$ 
and its derivatives) on the one hand, and lighter species 
(like H$_2$, N$_2$) on the other.  On the basis of our 
experiments, we identify CS$_2$ as the molecule 
that undergoes dynamic alignment when irradiated by 
long (35-ps) pulses; on the other hand, 100-fs pulses 
give rise to anisotropic fragment ion distributions that 
can be accounted for in terms to geometric alignment.

In our femtosecond experiments, light pulses (of 
wavelength 806 nm) were obtained from a high-intensity, 
Ti:S, chirped pulse amplified, 
100-fs laser operating at 10 Hz repetition rate. The laser 
light was focused using a biconvex lens, of focal length
10 cm, in an ultrahigh vacuum chamber capable of being 
pumped down to a base pressure of 3$\times$10$^{-11}$ 
Torr. We used operating pressures 
of $\leq$8$\times$10$^{-8}$ Torr (i.e. well below the 
pressure at which space charge effects will alter the 
results). Ions produced in the laser-molecule interaction 
zone were analyzed by a two-field, linear time-of-flight 
(TOF) spectrometer. To study the spatial distribution of ions 
produced in the focal zone, 
apertures of different sizes were inserted before the detector
in order to spatially limit the interaction volume being 
sampled. In the present series of measurements we 
used circular apertures, of  2 mm and 15 mm diameter, 
centered about the focal point. In the former case only 
the Rayleigh range (2.4 mm) was sampled while in the 
latter instance the lowest intensity accessed was 
5$\times$10$^{12}$ W cm$^{-2}$ for a 
peak laser intensity of 2$\times$10$^{15}$ W cm$^{-2}$.
Details of our apparatus and methodology are
presented elsewhere \cite{pramana}. Our picosecond 
experiments used the second harmonic from an Nd:YAG 
laser producing 35-ps long light pulses. Here, the ions 
formed were analyzed by either a 
quadrupole mass spectrometer or a TOF device. This 
apparatus has also been described in a number of 
earlier publications \cite{ourwork}.

Typical angular distributions measured for S$^+$ and S$^{2+}$ 
fragments are shown in Fig.1. The polarization angle was varied 
by means of a halfwave plate, with on-line monitoring of the laser 
intensity to ensure a constant value in the course of measurements 
with different polarizations. The angular distributions for 
S$^+$ and S$^{2+}$ (and for higher charge states of S-ions that 
are not shown in the figure)
are clearly very anisotropic, with many more ions being produced
in the direction of the laser polarization vector than in an 
orthogonal direction. This holds for both
35-ps and 100-fs duration laser pulses. {\it A priori}, it is not 
possible to deduce whether the observed anisotropy is due to 
dynamic or geometric effects.
Following the prescription articulated by Posthumus {\it et al.}
\cite{posthumus}, we distinguish between dynamic alignment
on the one hand and the effects of angle-dependent ionization rates 
(geometric alignment) on the other by probing the ratio of fragment 
ion yields obtained with orthogonal laser polarizations
over a range of laser intensities. Fig. 2 depicts the variation with
laser intensity of the ratio
of S$^{~+}_{\parallel}$/S$^{~+}_{\perp}$ (and the corresponding ratio 
for S$^{2+}$ and S$^{3+}$ ions), where the subscripts $\parallel$ and 
$\perp$ denote, respectively, the S$^+$ yield
at angles of 0$^\circ$ and 90$^\circ$ between the laser polarization 
vector and the axis of the TOF spectrometer. In 
the case of geometric alignment, it would be expected that the 
$\perp$-component becomes enhanced
as the laser intensity is increased. Consequently, the 
S$^{~+}_{\parallel}$/S$^{~+}_{\perp}$ ratio would fall with increasing 
laser intensity. Our 100-fs results indeed indicate this:
significant falls occur in the S$^{~+}_{\parallel}$/S$^{~+}_{\perp}$, 
S$^{~2+}_{\parallel}$/S$^{~2+}_{\perp}$ and  
S$^{~3+}_{\parallel}$/S$^{~3+}_{\perp}$ ratios as the laser intensity 
is increased from 10$^{13}$ to 10$^{15}$ W cm$^{-2}$. 
Geometric alignment clearly dominates the spatial alignment 
process in this case. However, Fig. 2 also shows that when 
35-ps duration laser pulses are used, of intensity in the 10$^{13}$ 
W cm$^{-2}$ range, the {\it opposite} effect is observed. The 
S$^{~+}_{\parallel}$/S$^{~+}_{\perp}$ 
ratio now increases with laser intensity. Similar 
observations were also made for S$^{2+}$ ions. Dynamic 
alignment of CS$_2$ clearly occurs when we use longer-duration
(35-ps) laser pulses. 

In order to gain some intuitive insight into the different behavior
obtained with short and long pulses, we have carried out 
calculations of the alignment dynamics by solving the classical 
equation of motion for a rigid rotor in an electric field (see (1)),
for different laser intensities and 
pulse durations. These calculations provide information on the
nature of the torques that are experienced by the molecule in the 
time evolution of the laser pulse. The interaction of the radiation 
field with CS$_2$ is, in the first approximation, governed 
by the molecular polarizability, 
$\alpha = \alpha_{\parallel} - \alpha_{\perp}$, where the first 
and second terms refer, respectively, to polarizability 
components parallel to and perpendicular to the molecular bond. 
Following Landau and Lifshitz \cite{landau} we express the 
angular acceleration of the internuclear axes as 
\begin{equation}
\ddot{\theta}= 
- \frac{\rm {\alpha}{\epsilon}^{2}}{2I}sin2{\theta},
\end{equation}
where $\theta$ is the polar angle between the S-C-S axis and 
the light field, $\epsilon$ is the field strength, and $I$ is the 
moment of inertia of the molecule about its centre of mass. 
We assume cylindrical symmetry and ignore higher-order 
terms involving $\alpha^2$. There is no 
permanent dipole contribution since we take CS$_{2}$ to remain 
linear even in a strong external field. 
The alignment dynamics calculated by us are depicted in Fig. 3
for a range of laser intensities.  The time-dependent light pulse is 
taken to be a gaussian multiplied by a cosine function (the 
intensity envelopes of our 100-fs and 35-ps pulses are shown as 
the solid lines in Fig. 3). As the
light intensity increases, a torque is exerted on the molecule, 
causing reorientation along the polarization vector. Further
increases in intensity lead to ionization, bond
stretching and multiple electron ejection (and consequent
dissociation). Since eqn. (1) only accounts for the first
of these steps, the reorientation time that is obtained is to
be regarded as a lower limit. Using 100-fs pulses at 
intensities $<$10$^{14}$ W cm$^{-2}$, our results indicate 
that no significant reorientation of the S-C-S molecule occurs. 
For peak intensities in the 10$^{14}$
W cm$^{-2}$ range, there is significant reorientation; the
angle changes from 0.75 to 0.45 radians in the time taken
for the laser pulse to reach an intensity of 
$\sim$5$\times$10$^{13}$ W cm$^{-2}$. Since this intensity
is well above the S$^+$ appearance threshold (Fig. 2), 
the extent of reorientation is clearly overestimated in
our calculations. For higher peak intensities
($\sim$10$^{15}$ W cm$^{-2}$), our results indicate
that although the torque generated by the laser field is
sufficient to bring the molecule in line with the polarization 
vector, the angular velocity imparted in the process is
large enough to cause oscillations about the polarization
axis such that there is no overall alignment. 
Fig. 3b shows the corresponding calculations for
35-ps laser pulses. For peak intensities above
10$^{12}$ W cm$^{-2}$, the molecular axis becomes 
coincident with the polarization vector very early in
the time evolution of the laser pulse. Thereafter, small
amplitude oscillations are seen. Such a system would
be expected to show dynamic alignment, as is indeed
seen in our experiments.  

An obvious, but presently unavoidable, shortcoming 
in our calculations is in our use of 
$\alpha$-values that pertain to CS$_2$ in its ground electronic 
state. The value of 
$\alpha_{\parallel}$ will increase substantially with applied field, as
the bond elongates and the electron density distribution is distorted
from its ground-state morphology. No information exists on how, and
to what extent, such enhancement occurs. In addition, contributions
from the hyperpolarizabilities (which are expected to be substantial
at these laser intensities) must also be considered in a proper
description \cite{hyper}. Nevertheless, the qualitative insight that
these model calculations yield is encouraging in that there is
consistency with our experimental observations. 

We note that the prescription used by Ellert and Corkum
\cite{ellert} in order to distinguish between dynamic and geometric
alignment in iodine and iodine-containing molecules was to measure
the fragmentation pattern using linearly- and circularly-polarized 
light of intensities such that the same field strengths were
obtained in both cases. Identical fragmentation patterns were taken
as evidence against dynamic alignment. An assumption that is
implicit in such an approach is that circularly-polarized light 
can be treated as a combination of two perpendicular, 
linearly-polarized components. However, on the basis of recent
experiments, it is our contention 
that the dynamics resulting from irradiation of molecules by 
circularly-polarized light cannot, {\it a priori}, be expected to be a 
linear combination of the dynamical effects due to linearly 
polarized light aligned parallel and perpendicular to the molecular 
symmetry axis. Circularly-polarized light imparts angular 
momentum to the molecule that is being irradiated, whereas 
linearly-polarized light does not. How this might affect molecular 
dynamics in intense laser fields is an issue that has not been 
properly addressed. Experiments that we have recently 
conducted on intense-field-induced multiple ionization of 
N$_2$ \cite{rapidcomm} 
reveal that the polarization state of the incident light affects 
the ionization spectrum in the following fashion: when using 
circularly-polarized light, we observe a distinct {\em 
suppression of ionization channels} compared to the situation 
pertaining to linearly-polarized light of the same field strength. 
Moreover, an enhancement of lower-energy pathways to 
dissociation is apparent in the case of circular polarization. 
We believe that this may reflect the 
importance of high angular momentum intermediate states 
that may be accessed when circularly-polarized light is used.   
Such states present ``wider" centrifugal barriers to 
dissociative ionization; this manifests itself in the increasing 
importance of longer tunneling time pathways that our 
data on N$_2^{~q+}$ ions \cite{rapidcomm} indicate.

In the case of CS$_2$ molecules also, we observe significant
differences in the pattern of dissociative ionization between 
circularly- and linearly-polarized light. By way of example, 
we show in Fig. 4 the fragmentation pattern obtained, at two 
laser intensities, using linearly-polarized light that is aligned
parallel and perpendicular to the TOF axis, as well as with 
circularly-polarized light. The ion yield obtained with circularly-
polarized light is uniformly lower than that obtained with linear 
polarization, parallel as well as perpendicular (the
CS$_2^{~+}$ peak at the higher laser intensity is not 
taken into account in this comparison as the ion signal was 
saturated). Note that fragment ion yields obtained with
parallel polarization are higher than corresponding yields
obtained with circular-polarization, even when the magnitudes
of the electric field components in the latter are a factor of
three larger than in the former.
Hence, comparison of ion yields obtained with linear
and circular polarization cannot give unambiguous evidence 
for or against dynamic alignment, and it is for this reason 
that we opt to rely on data shown in Fig. 2 to make deductions
about geometric alignment being responsible for the anisotropic
angular distributions that are obtained for fragment ions when 
CS$_2$ molecules are immersed in 100-fs-long laser pulses. 

Hitherto, discussions of polarization effects in molecules have 
tended to focus only on classical aspects of spatial 
alignment resulting from induced dipole moments in intense 
light fields. The results shown in Fig. 4 indicate that the polarization 
state of light is also of fundamental importance in a quantum 
mechanical sense in that it affects molecular ionization yields 
and dissociation pathways. It is clear that such 
considerations need to be incorporated in development 
of adequate descriptions of molecular dynamics in 
intense light fields.

We thank the Department of Science and Technology 
for substantial financial support for our femtosecond laser system
and Vinod Kumarappan for many useful discussions.

\begin{figure}
\caption{Variation of (a) S$^+$ and (b) S$^{2+}$ yield with polarization 
angle. These measurements were made using 100-fs laser pulses. 
Similar data was obtained using 35-ps pulses.}
\end{figure}

\begin{figure}
\caption{Variation with laser intensity of the ratio,
S$^{~q+}_{\parallel}$/S$^{~q+}_{\perp}$. 
$\parallel$ and $\perp$ denote, respectively, the ion yield
at angles of 0$^\circ$ and 90$^\circ$ between the laser polarization 
vector and the TOF axis. Note the different functional 
dependence obtained in the case of 35-ps and 100-fs pulses.
The picosecond ratios have been divided by 2.}
\end{figure}

\begin{figure}
\caption{Calculated time dependence of the angle between the S-C-S axis
and the laser polarization vector for different values of laser intensity for (a)
100-fs and (b) 35-ps pulses. The temporal
profiles of the laser pulses are indicated by the solid curves.}
\end{figure}

\begin{figure}
\caption{Mass spectra of CS$_2$ irradiated by 100-fs pulses at two laser
intensities, with linear polarization (pointing parallel and perpendicular
to the TOF axis) as well as with circular polarization.}
\end{figure} 

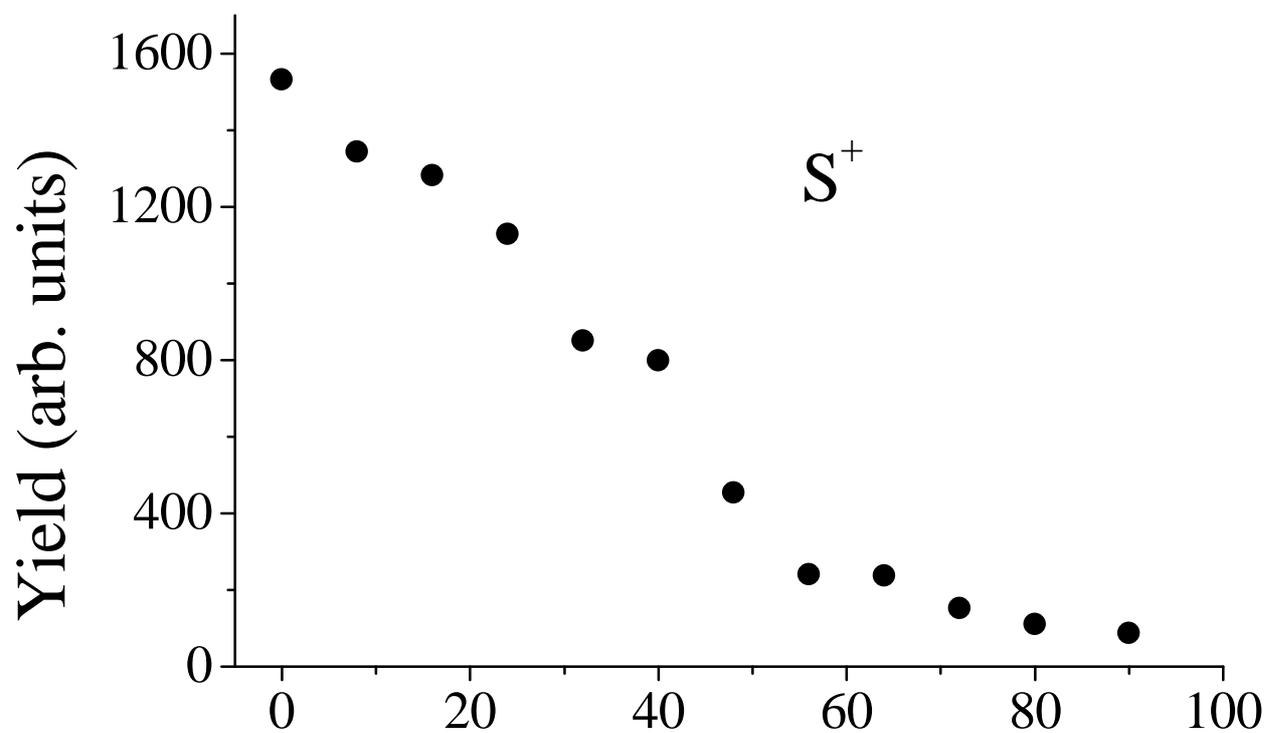
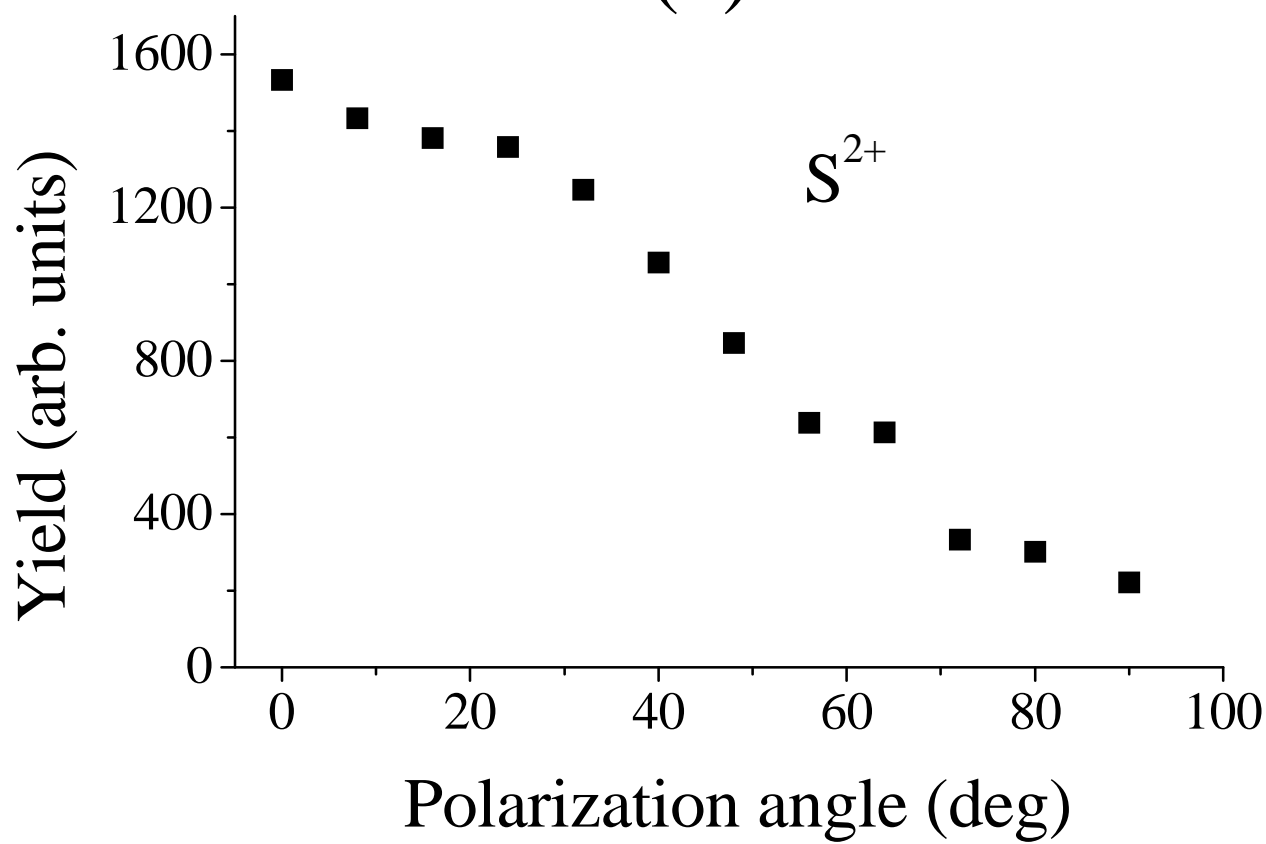
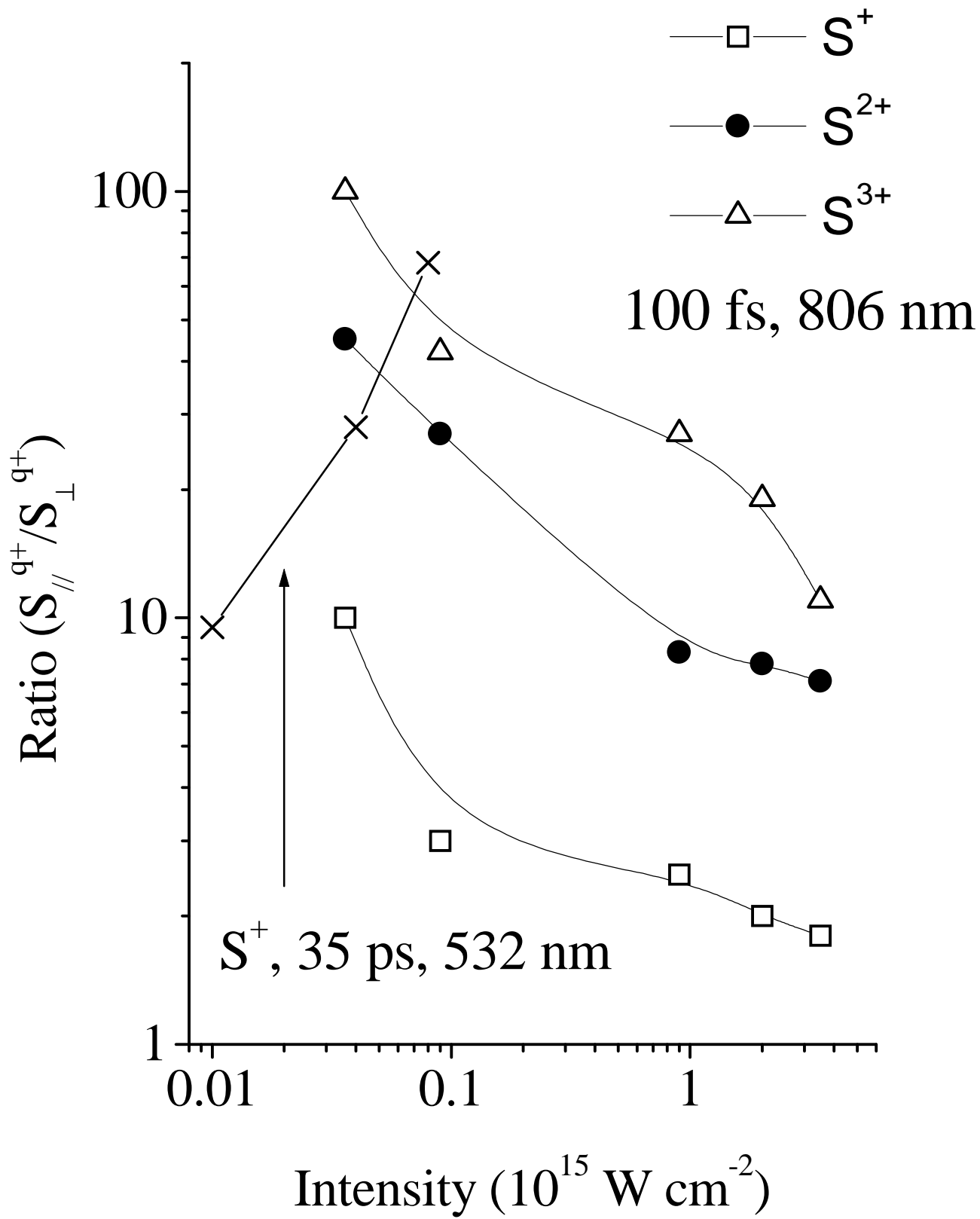
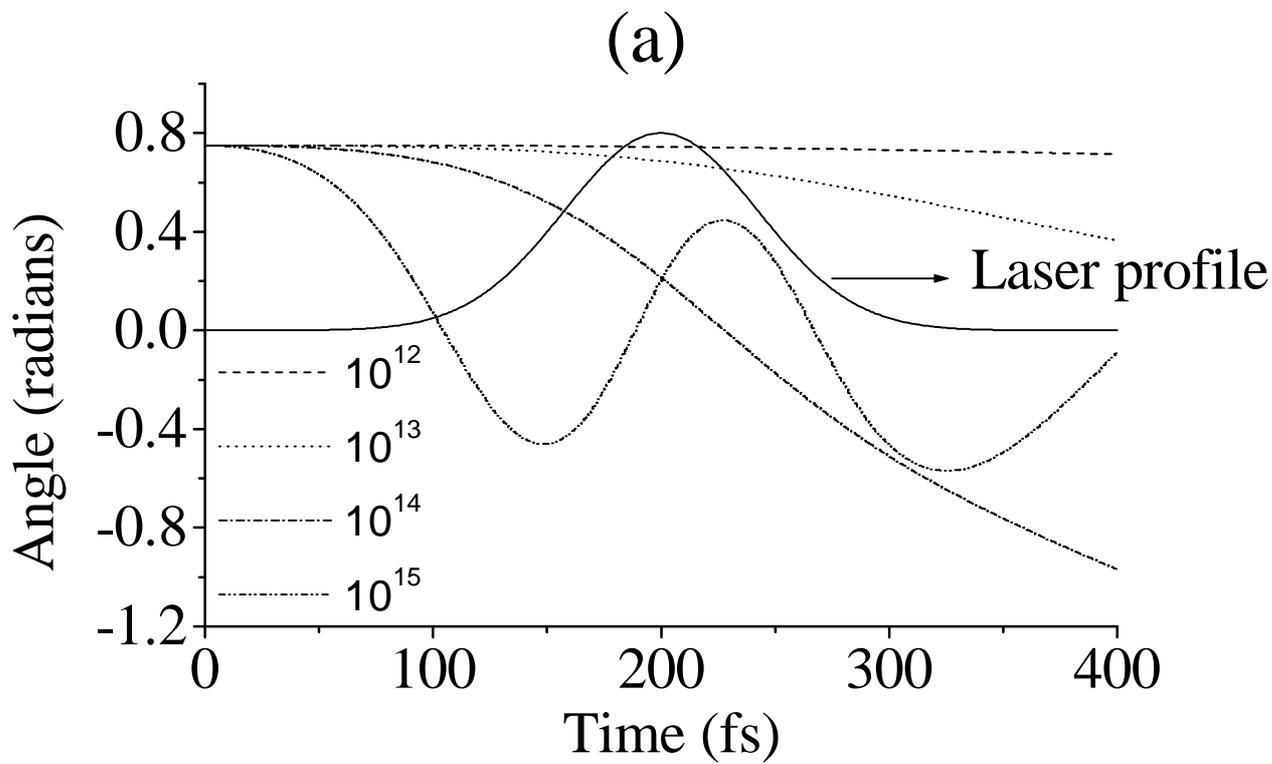
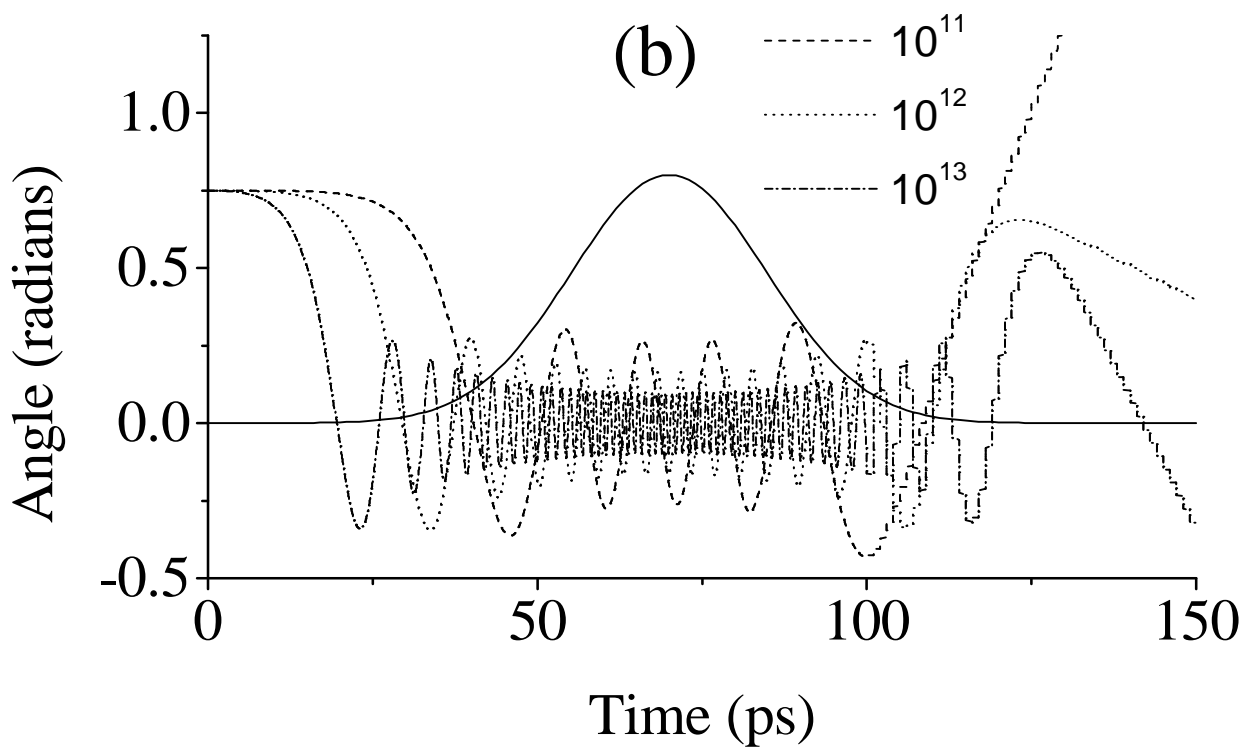
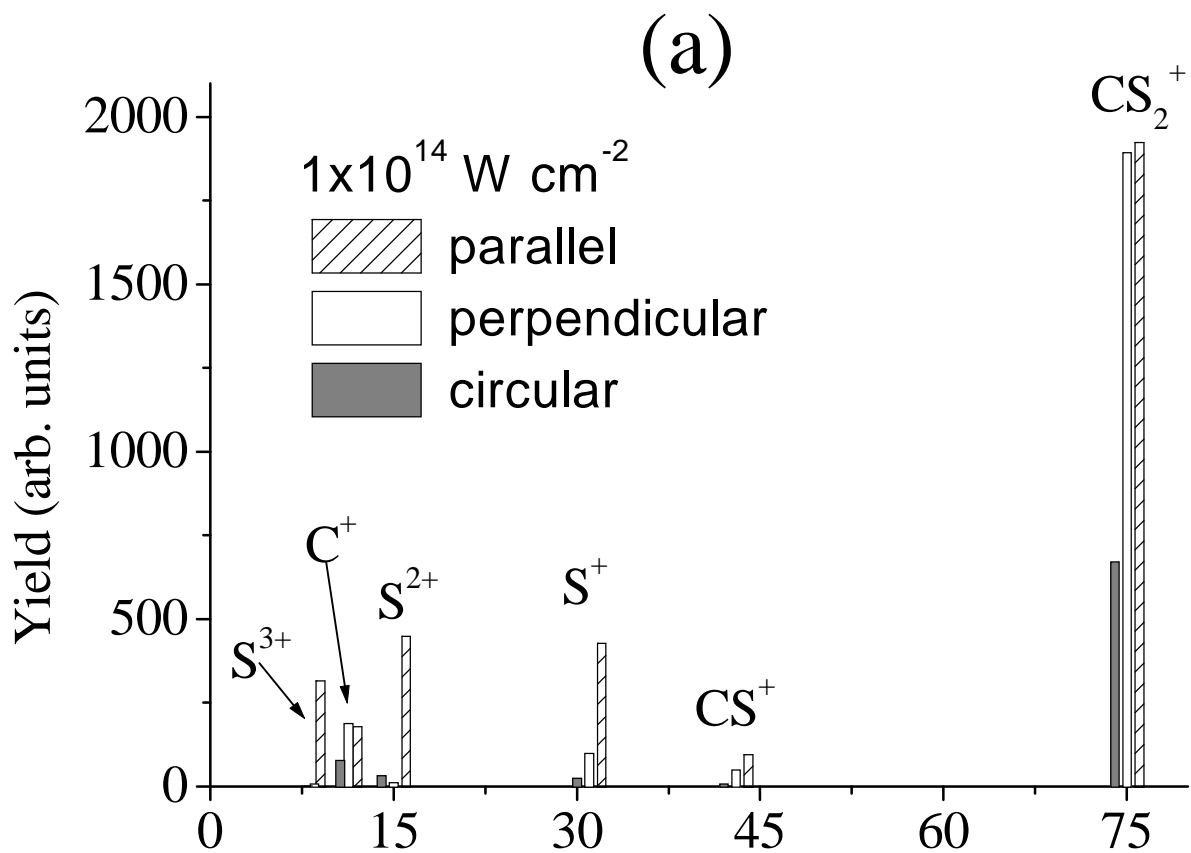
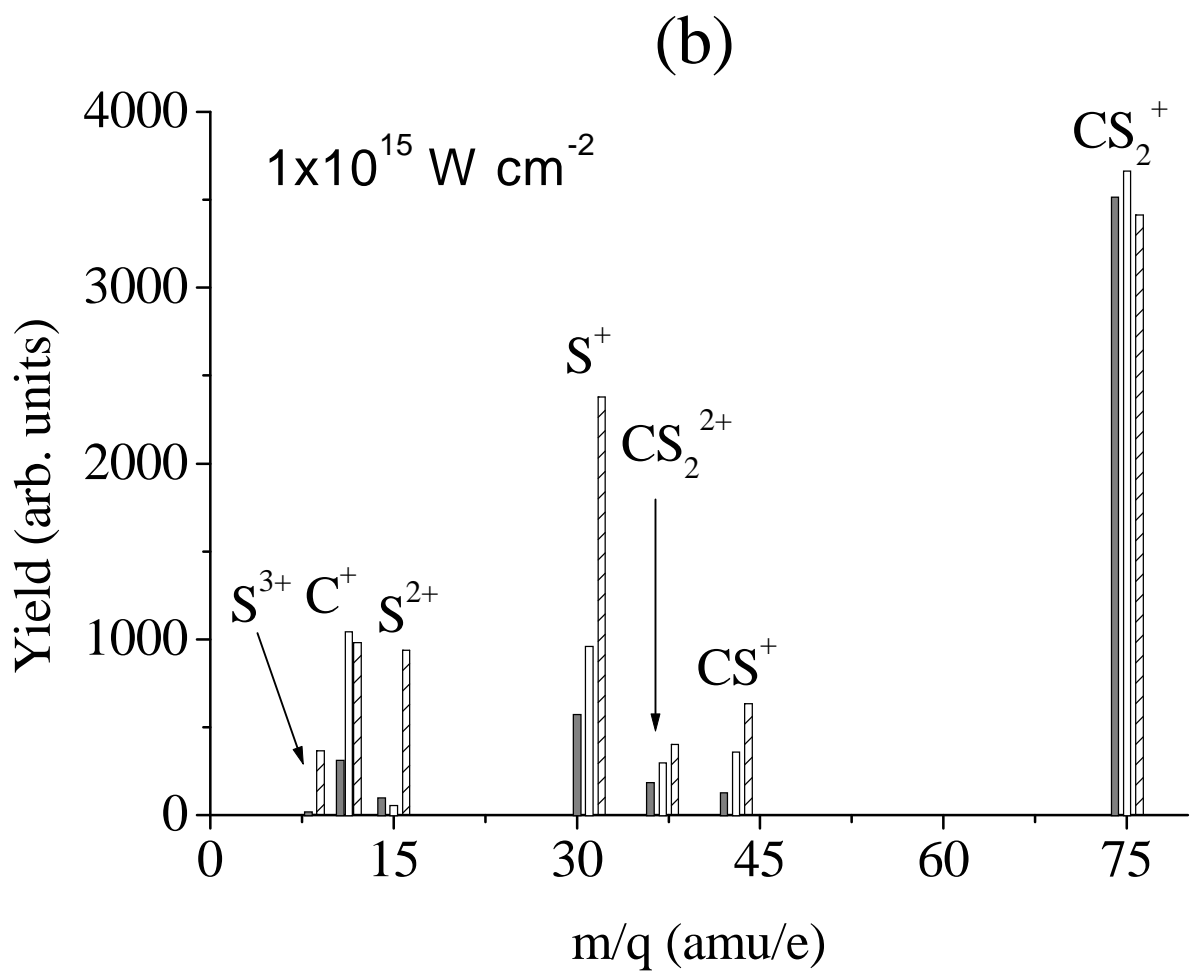
\end{document}